\documentclass[twocolumn,aps,prd,preprintnumbers,showpacs,showkeys,nofootinbib,superscriptaddress,fleqn,floatfix,tightenlines,10pt]{revtex4-1}
\usepackage{amsmath,amsfonts,amssymb,amscd,amsxtra,amsthm}
\usepackage{graphicx}
\usepackage{epstopdf}
\usepackage{bm}
\usepackage{hyperref}
\usepackage[normalem]{ulem} 
\usepackage[dvipsnames]{xcolor} 
\renewcommand\sout{\bgroup \color{red} \ULdepth=-.5ex \ULset}

\usepackage[section]{placeins}
\usepackage[caption = false]{subfig}
\usepackage{multirow}
\usepackage{cmupint}
\newcommand{\be}{\begin{eqnarray}}
\newcommand{\ee}{\end{eqnarray}}

\begin{document}
\preprint{}
\title{Description of nucleon transfer reactions at intermediate energies within the impulse picture}
\author{Sang-In Shim}
\email[E-mail: ]{shimsang@rcnp.osaka-u.ac.jp}
\affiliation{Research Center for Nuclear Physics (RCNP),
Osaka University, Ibaraki, Osaka, 567-0047, Japan}
\affiliation{Center for Extreme Nuclear Matters (CENuM),
Korea University, Seoul 02841, Republic of Korea}
\author{Yoshiki Chazono}
\email[Present address: ]{Department of Physics, Kyushu University, Fukuoka 819-0395, Japan}
\affiliation{RIKEN Nishina Center for Accelerator-Based Science, 
2-1 Hirosawa, Wako 351-0198, Japan}
\author{Kazuki Yoshida}
\email[Present address: ]{Research Center for Nuclear Physics (RCNP), Osaka University, Ibaraki, Osaka, 567-0047, Japan; \\
Interdisciplinary Theoretical and Mathematical Sciences Program (iTHEMS), RIKEN, Wako 351-0198, Japan}
\affiliation{Advanced Science Research Center, 
Japan Atomic Energy Agency, Tokai, Ibaraki 319-1195, Japan}
\author{Tomohiro Uesaka}
\affiliation{RIKEN Nishina Center for Accelerator-Based Science, 
2-1 Hirosawa, Wako 351-0198, Japan}
\author{Kazuyuki Ogata}
\affiliation{Department of Physics, Kyushu University, 
Fukuoka 819-0395, Japan}
\affiliation{Research Center for Nuclear Physics (RCNP),
Osaka University, Ibaraki, Osaka, 567-0047, Japan}
\date{\today}
\begin{abstract}
\noindent
 {\bf Background}:
 At intermediate energies, transfer reactions are suppressed because the momentum-matching condition is difficult to satisfy. In the standard distorted wave Born approximation (DWBA), a high momentum component of the transferred particle is required to match the large momentum transfer.
 \\
 {\bf Purpose}:
 We investigate the applicability of the distorted wave impulse approximation (DWIA) for describing ($p,d$) transfer reactions at intermediate energies by performing a comparative study with the standard DWBA. DWIA, which has been successful for knockout reactions, is expected to provide an alternative reaction mechanism at this energy region.
 \\
 {\bf Methods}:
 Both DWBA and DWIA formalisms are applied to the $^{16}$O($p,d$){}$^{15}$O reaction at 200~MeV. In DWBA, the reaction is described as a neutron pickup, while in DWIA, it is treated as a quasi-elastic scattering from a preformed deuteron cluster in the target.
 \\
 {\bf Results}:
 The DWBA calculation is in good agreement with the experimental data, reproducing both the angular distribution and the absolute magnitude of the cross section with a reasonable spectroscopic factor. In contrast, the DWIA calculation, while qualitatively reproducing the trend of the angular distribution, severely underestimates the cross section by about two orders of magnitude.
 \\
 {\bf Conclusions}:
 Our findings suggest that conventional DWBA provides a more suitable description for the $^{16}$O($p,d$){}$^{15}$O reaction at 200~MeV. The failure of DWIA in this case, unlike its success in knockout reactions, raises open questions about its applicability to transfer reactions. This motivates the need for systematic investigations to delineate the applicability of both reaction mechanisms under various conditions.
\end{abstract}
\pacs{}
\keywords{}
\maketitle
\section{Introduction}
\label{sec:intro}
Proton-induced neutron pickup ($p,d$) reactions have been used to investigate the neutron single particle (s.p.) structure of atomic nuclei. 
To realize the pickup reactions, the momentum-matching condition must be satisfied, that is, the momentum of the neutron to be picked up must be similar to the momentum transfer. 
Because the motion of a nucleon inside a nucleus is restricted by the Fermi momentum, the momentum-matching condition becomes difficult to satisfy at higher incident energies. 
Quite recently, Ong and collaborators claimed that the ($p,d$) reaction at intermediate energies can be a good probe for a high-momentum neutron in a nucleus~\cite{Ong:2012lwd}, despite the small cross section due to the breakdown of the momentum-matching condition. 
In such situations, where the reaction dynamics can be more complex, it is worthwhile to consider an alternative approach to the standard distorted wave Born approximation (DWBA).

In this work, we propose to apply the distorted wave impulse approximation (DWIA)~\cite{Chant:1977zz,Chant:1983zz} to the description of ($p,d$) reactions at intermediate energies, and compare the result with that of the conventional DWBA.
DWIA is a standard reaction model for knockout reactions. 
In the picture of DWIA, the incident particle collides only with a particle in a nucleus and knocks it out of the nucleus without affecting the rest part of the nucleus. 
DWIA has successfully been applied to various knockout reactions such as ($p,p$N)~\cite{Wakasa:2017rsk}, ($p,p\alpha$)~\cite{Yoshida:2019cxw,Taniguchi:2021ulj}, and $(p,pd)$~\cite{Samanta:1982xk,Terashima:2018bwq}. 
Among them, Ref.~\cite{Terashima:2018bwq} aimed to extract information on the high momentum component of a neutron due to the tensor correlation, as in Ref.~\cite{Ong:2012lwd}. 
They chose kinematics of ($p,pd$) so that the elementary process corresponded to the $p$-$d$ backward angle scattering to pin down a $pn$ pair having high relative momentum in a nucleus. 
Because the $p$-$d$ scattering at backward angles is a pickup-type process, the ($p,pd$) process studied in Ref.~\cite{Terashima:2018bwq} can be understood that the incident proton picks up a neutron that is strongly correlated with a proton in a nucleus. 
Then, after the $p$-$d$ process, the proton that formed a $pn$ pair goes outside the nucleus. 
Here, if we consider a process in which the proton is bound in the nucleus in the final channel, it is nothing but the ($p,d$) transfer reaction. 
This suggests that the transfer reaction can also be described within a DWIA-like framework, which fundamentally assumes a quasi-elastic scattering process with a preformed deuteron cluster inside the nucleus, in contrast to the neutron pickup picture of DWBA.
This approach is particularly compelling for intermediate energies, as this is the regime where DWIA is well-established for describing knockout reactions with high momentum transfer~\cite{Wakasa:2017rsk, pty011_etal, Noro:2020mtr, Noro:ptad116}.
As discussed in detail below, this DWIA framework has the advantage of making it easier to satisfy the momentum-matching condition at intermediate energy.

To test this approach, we apply the DWIA formalism, alongside a standard DWBA calculation, to the $^{16}$O($p,d$){}$^{15}$O reaction at the incident energy of 200 MeV.
This study expands upon the foundational concept explored in Refs.~\cite{Kam00,Ues02} by one of the authors of this paper and collaborators, which was based on a similar idea but restricted to forward-angle kinematics.
The primary goal of this paper is to directly compare the results from both the DWIA and DWBA calculations against experimental data, thereby assessing the strengths and limitations of each approach for describing the ($p,d$) process at intermediate energies.

The content of this paper is as follows.
In Section~\ref{sec:formalism}, we first briefly outline the standard DWBA formalism for transfer reactions and then present the detailed derivation of the DWIA framework. 
In Section~\ref{sec:result}, we show the numerical results of both calculations for the $^{16}$O($p,d$){}$^{15}$O reaction and compare them with experimental data. 
We also discuss the different effects of distortion in the two models.
Section~\ref{sec:summary} is devoted to a summary and conclusion.

\section{Formalism}
\label{sec:formalism}
\subsection{The distorted wave Born approximation}
\label{subsec:DWBA}
In this subsection, we briefly review the standard formalism of DWBA for the A($p,d$)B transfer reaction. 
Within a three-body model consisting of a proton ($p$), a neutron ($n$), and a core nucleus (B), the transition amplitude for the A($p,d$)B reaction is given by the prior-form DWBA transition matrix~\cite{Pang:2014kha,Xu:2024ohi}
\begin{equation} 
  T_{pd}^{\text{DWBA}} 
  = S_{nlj}^{1/2} \langle \chi^{(-)}_{\bm{K}_{\rm f}} \phi^{(d)} | V_{np} | \chi^{(+)}_{\bm{K}_{\rm i}} \psi_{nlj} \rangle, 
  \label{eq:T_DWBA_full} 
\end{equation} 
where $S_{nlj}^{1/2}$ is the spectroscopic amplitude for the neutron in the target nucleus A (= B + $n$). 
The quantum numbers $n$, $l$, and $j$ denote the radial quantum number, orbital angular momentum, and total angular momentum, respectively, of the neutron s.p. wave function, $\psi_{nlj}$. 
$\chi^{(+)}_{\bm{K}_{\rm i}}$ and $\chi^{(-)}_{\bm{K}_{\rm f}}$ are the distorted waves for the $p$-A and $d$-B systems, respectively, where $\bm{K}_{\rm i}$ and $\bm{K}_{\rm f}$ are the momenta of the incident proton and the emitted deuteron, and the superscripts $(+)$ and $(-)$ denote the outgoing and incoming boundary conditions.
$V_{np}$ is the interaction potential between the proton and neutron that forms the deuteron, whose wave function is $\phi^{(d)}$.

To simplify the calculation of the matrix element in Eq.~\eqref{eq:T_DWBA_full}, the zero-range approximation is often employed~\cite{SATCHLER19641,Johnson:1970bh,Ohm70}.
This approximation can be expressed as 
\begin{equation} 
  V_{np}({\bm r}_{np})\phi^{(d)}({\bm r}_{np}) \approx D_0 \delta({\bm r}_{np}),
  \label{eq:zero_range_approx} 
\end{equation} 
where ${\bm r}_{np} = {\bm r}_n - {\bm r}_p$ is the relative coordinate between the neutron and the proton. $D_0$ is a constant representing the strength of the interaction, which is determined from the integrated value of Eq.~\eqref{eq:zero_range_approx}.

By applying this approximation to Eq.~\eqref{eq:T_DWBA_full}, the transition matrix in DWBA is simplified to
\begin{equation} 
  T_{pd}^{\text{DWBA}} 
  \approx S_{nlj}^{1/2} D_0 \int d {\bm r'} \, \chi^{(-)*}_{\bm{K}_{\rm f}}({\bm r'}) \psi_{nlj}({\bm r'}) \chi^{(+)}_{\bm{K}_{\rm i} }(\mathbf{r'}). 
  \label{eq:T_DWBA_zero_range} 
\end{equation} 

\subsection{The distorted wave impulse approximation}
Let us consider the A$(p, d)$B reaction as in Sec.~\ref{subsec:DWBA}. Now we describe this process as a collision between the incident proton and a deuteron in A. The nuclear wave function in the initial state is thus given by a $d+{\rm C}$ two-body bound state:
\begin{align}
  \Phi_{\mathrm A}= & S_d^{1/2} \psi^{(d)}({\bm R}_2) \phi^{(d)}({\bm r}) \sum_{\mu_d} \frac{(-)^{1-\mu_d}}{\sqrt{3}} \eta^{(d)}_{1\mu_d} \Phi_{{\mathrm C},-\mu_d}, 
  \label{eq:wfA0}
\end{align}
where $S_d^{1/2}$ is the deuteron spectroscopic amplitude, $\psi^{(d)}$ ($\phi^{(d)}$) is the $d$-C ($p$-$n$) relative wave function, $\eta^{(d)}_{1\mu_d}$ is the spin 1 wave function with $\mu_d$ being the spin third component, and $\Phi_{{\mathrm C},-\mu_d}$ is the wave function of C. Here, we have assumed that the spin-parity of A (C) is $0^+$ ($1^+$) and the orbital angular momentum between $d$ and C is zero; the third component of the total spin 1 of C is $-\mu_d$ because of the angular momentum conservation. The wave function of B is given by
\begin{align}
  \Phi_{\mathrm B} = & S_p^{1/2}
   \varphi^{(p)}_{nlj}(R_1)\sum_{\mu_{\mathrm C}\mu_j} 
   (j \mu_j 1 \mu_{\mathrm C} | I_{\mathrm B} \mu_{\mathrm B}) \nonumber \\
   & \times
   \left[ Y_l(\bm{ \hat{R}_1 }) \otimes \eta^{(p)}_{1/2} \right]_{j\mu_j}
   \Phi_{{\mathrm C}\mu_{\rm C}},
   \label{eq:wfB0}
\end{align}
where $S_p^{1/2}$ is the proton spectroscopic amplitude for which C is in the $1^+$ state, $\varphi^{(p)}_{nlj}$ is the $p$-C relative wave function; $n$, $l$, and $j$ are the same as those in Sec.~\ref{subsec:DWBA}, but for $p$ with respect to C.
\begin{figure}[htp]
  \includegraphics[width=8cm]{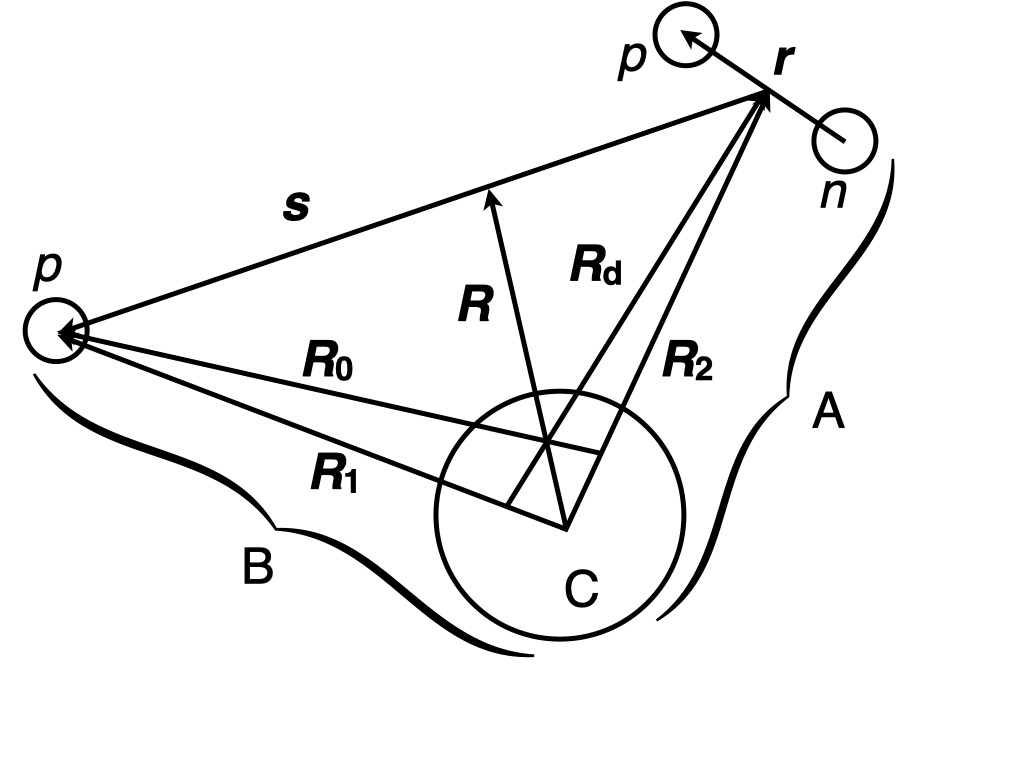}
  \caption{Coordinates of the A($p$, $d$)B reaction system.}
  \label{fig:1}
  \end{figure}

In the impulse picture, C is regarded as a spectator. Therefore, the nuclear wave function appearing in the transition matrix is
\begin{align}
  \int \Phi_{\rm B}^* \Phi_{\rm A} d\xi_{\rm C}=&
  S_p^{1/2} S_d^{1/2} \varphi^{(p)}_{nlj}(R_1) \psi^{(d)}({\bm R_2}) \phi^{(d)}({\bm r}) \nonumber \\
  & \times \sum_{\mu_j \mu_d} (j \mu_j 1, -\mu_d | I_{\mathrm B} \mu_{\mathrm B}) \frac{(-)^{1-\mu_d}}{\sqrt{3}} \nonumber \\
  & \times \eta^{(d)}_{1\mu_d} \left[ Y_l(\bm{ \hat{R}_1 }) \otimes \eta^{(p)}_{1/2} \right]_{j\mu_j}^{*},
  \label{eq:overlabAB2}
\end{align}
where $\xi_{\rm C}$ denotes the internal coordinates of C. The transition matrix of the  A$(p, d)$B reaction with the impulse picture is given by
\begin{eqnarray}
  T^{(pd)}_{\mu'_d \mu_{\rm B} \mu_0}
  &=&S_p^{1/2} S_d^{1/2} \sum_{\mu_j \mu_d} (j \mu_j 1, -\mu_d | I_{\mathrm B} \mu_{\mathrm B}) \frac{(-1)^{1-\mu_d}}{\sqrt{3}}
  \nonumber\\
  &&\times\sum_{m \mu_{N}}
  \left(l m \frac{1}{2} \mu_N \bigg| j \mu_j \right)
  \bar{T}^{(pd)}_{m,\mu'_d \mu_N \mu_0 \mu_d},
  \label{eq:Tamp}
\end{eqnarray}
where
\begin{eqnarray}
  \bar{T}^{(pd)}_{m,\mu'_d \mu_N \mu_0 \mu_d}
  & \equiv &
  \biggl \langle
  \chi^{(-)}_{\bm{K}_{\rm f}}(\bm{R}_{d} )
  \phi^{(d)}(\bm{r}) \eta^{(d)}_{1\mu'_d}
  \psi^{(p)}_{nljm}(\bm{R}_{1}) \eta^{(p)}_{\frac{1}{2}\mu_N} \biggl| \nonumber\\
  &&\times t_{pd}(\bm{s}) \nonumber \\
  &&\times\biggr| \chi^{(+)}_{\bm{K}_{\rm i}}(\bm{R}_{0})\eta^{(p)}_{\frac{1}{2}\mu_0}
  \phi^{(d)}(\bm{r} ) \eta^{(d)}_{1\mu_d}
  \psi^{(d)}(\bm{R}_{2})
  \biggl \rangle. \nonumber \\
  \label{eq:barTamp}
\end{eqnarray}
Here, the $p$-C relative wave function is denoted by
\begin{align}
  \psi^{(p)}_{nljm}({\bm R}_1)=\varphi^{(p)}_{nlj}(R_1)  Y_{lm}(\bm{ \hat{R}_1 })
\end{align}
and $t_{pd}$ is an effective interaction between $p$ and $d$.

The coordinates $\bm{R}_1$, $\bm{R}_2$, $\bm{R}_d$, and $\bm{R}_0$ can be written with the $p$-$d$ relative coordinate $\bm{s}$ and the center-of-mass (c.m.) coordinate $\bm{R}$ of the $p$-$d$ system with respect to C as
\begin{align}
  \bm{R}_1 &= \bm{R} + \frac{2}{3}\bm{s}, \\
  \bm{R}_2 &= \bm{R} - \frac{1}{3}\bm{s}, \\
  \bm{R}_d &=  \frac{B-1}{B}\bm{R} - \frac{2+B}{3B}\bm{s} \equiv \alpha_1 \bm{R}+\beta_1 \bm{s},\\
  \bm{R}_0 &=  \frac{A-2}{A}\bm{R} + \frac{2(A+1)}{3A}\bm{s} \equiv \alpha_2 \bm{R}+\beta_2 \bm{s},
  \label{eq:coord2}
\end{align}
where $A$ and $B$ are the mass numbers of A and B, respectively. We then employ the asymptotic momentum approximation (AMA)~\cite{Wakasa:2017rsk,yoshida16} to describe the propagation of the scattering wave for a short distance by a plane wave as follows:
\begin{align}
 \chi^{(-)}_{\bm{K}_{\rm f}}(\bm{R}_d) 
 &\approx \chi^{(-)}_{\bm{K}_{\rm f}}\left(\alpha_1\bm{R}\right)
 e^{i \bm{K}_{\rm f} \cdot \beta_1\bm{s} },
 \label{eq:AMAdwaves1} \\
 \chi^{(+)}_{\bm{K}_{\rm i}}(\bm{R}_0) 
 &\approx \chi^{(+)}_{\bm{K}_{\rm i}}\left(\alpha_2\bm{R}\right)
 e^{i \bm{K}_{\rm i} \cdot \beta_2 \bm{s} }.
 \label{eq:AMAdwaves2}
\end{align}
For the bound state wave functions, we use their Fourier transforms:
\begin{align}
  \psi^{(p)}_{nljm}(\bm{R}_1) &= \frac{1}{(2\pi)^3} \int \tilde{\psi}^{(p)}_{nljm}(\bm{k}_p) 
    e^{i \bm{k}_p \cdot \left( \bm{R} + \frac{2}{3}\bm{s} \right)}d\bm{k}_p,
    \label{eq:Boundwaves1} \\
  \psi^{(d)}(\bm{R}_2) &= \frac{1}{(2\pi)^3} \int \tilde{\psi}^{(d)}(\bm{k}_d) 
    e^{i \bm{k}_d \cdot \left( \bm{R} - \frac{1}{3}\bm{s} \right)}d\bm{k}_d.
  \label{eq:Boundwaves2}
\end{align}

Substituting Eqs.~\eqref{eq:AMAdwaves1}-\eqref{eq:Boundwaves2}, one finds
\begin{eqnarray}
  \bar{T}^{(pd)}_{m,\mu'_d \mu_N \mu_0 \mu_d}
  &\approx& \frac{1}{(2\pi)^6}
    \int d{\bm k_{d}} d{\bm k_{p}} \,
    \tilde{\psi}^{(p)*}_{nljm}(\bm{k}_{p}) \,
    \tilde{\psi}^{(d)}(\bm{k}_{d}) \nonumber \\
    &&\times \tilde{t}^{(pd)}_{\bm{\kappa}' \mu'_d \mu_N, \bm{\kappa} \mu_0 \mu_d}
    \int
    \chi^{(-)*}_{\bm{K}_{\rm f}}
    (\alpha_1 \bm{R})
    \nonumber \\
    &&\times
    \chi^{(+)}_{\bm{K}_{\rm i}}(\alpha_2 \bm{R} )e^{-i \bm{k}_p \cdot \bm{R}}
    e^{i \bm{k}_d \cdot \bm{R}}d{\bm R},
  \label{eq:barTamp2}
\end{eqnarray}
where $\tilde{t}^{(pd)}_{\bm{\kappa}' \mu'_d \mu_N, \bm{\kappa} \mu_0 \mu_d}$ is defined by
\begin{eqnarray}
  \tilde{t}^{(pd)}_{\bm{\kappa}' \mu'_d \mu_N, \bm{\kappa} \mu_0 \mu_d}
  & \equiv &
  \biggl \langle
   e^{i \bm{\kappa}' \cdot \bm{s} }\phi^{(d)}(\bm{r})\eta^{(d)}_{1\mu'_d}\eta^{(p)}_{\frac{1}{2}\mu_N} \biggl|  \nonumber\\
  &&\times t_{pd}(\bm{s}) \biggr| e^{i \bm{\kappa} \cdot \bm{s} }\phi^{(d)}(\bm{r}) \eta^{(p)}_{\frac{1}{2}\mu_0}\eta^{(d)}_{1\mu_d}
  \biggl \rangle. \nonumber \\
  \label{eq:tildetpd}
\end{eqnarray}
with
\begin{align}
 \bm{\kappa}'
  &=\frac{2}{3}\bm{k}_p - \beta_1\bm{K}_{\rm f},
  \label{eq:kappap}
  \\
 \bm{\kappa}
  &=\beta_2\bm{K}_{\rm i} - \frac{1}{3}\bm{k}_d.
  \label{eq:kappas}
\end{align}
In the plane wave (PW) limit, one can find that $\bm{k}_p$ becomes as follows:
\begin{align}
\bm{k}_p \rightarrow \alpha_2\bm{K}_{\rm i} + \bm{k}_d -\alpha_1\bm{K}_{\rm f}.
\end{align}
We also use this property in the description of the $p$-$d$ process when distortions exist. Further, we assume $A\gg 1$ and $B\gg 1$ and simplify Eqs.~\eqref{eq:kappap} and \eqref{eq:kappas} as
\begin{align}
 \bm{\kappa}'
  &\approx \frac{2}{3}\bm{k}_p^{(0)} - \frac{1}{3}\bm{K}_{\rm f},
  \\
 \bm{\kappa}
  &\approx \frac{2}{3}\bm{K}_{\rm i} - \frac{1}{3}\bm{k}_d
\end{align}
with
\begin{align}
\bm{k}_p^{(0)} \equiv \bm{K}_{\rm i} + \bm{k}_d -\bm{K}_{\rm f}.
\label{eq:kp0}
\end{align}
One can find from Eq.~\eqref{eq:kp0} that the total momentum of the $p$-$d$ system is conserved. With this treatment, Eq.~\eqref{eq:barTamp2} becomes
\begin{eqnarray}
  \bar{T}^{(pd)}_{m,\mu'_d \mu_N \mu_0 \mu_d}
  &\approx& \frac{1}{(2\pi)^3}
    \int d{\bm k_{d}} \,
    \tilde{\psi}^{(d)}(\bm{k}_{d})
    \tilde{t}^{(pd)}_{\bm{\kappa}' \mu'_d \mu_N, \bm{\kappa} \mu_0 \mu_d}
    \nonumber \\
    &&\times
    \int \chi^{(-)*}_{\bm{K}_{\rm f}} (\bm{R})
    \chi^{(+)}_{\bm{K}_{\rm i}}(\bm{R} )\psi^{(p)*}_{nljm}(\bm{R})
    e^{i \bm{k}_d \cdot \bm{R}}d{\bm R}. \nonumber \\
  \label{eq:barTamp3}
\end{eqnarray}

The differential cross section for the A$(p,d)$B transfer reaction is given by
\begin{align}
  \frac{d\sigma_{pd}}{d\Omega}
  &=\frac{C_0}{6}S_{p}S_{d}\nonumber\\
  &\times\sum_{\mu'_d \mu_{\mathrm B} \mu_0}
   \left| \sum_{\mu_j \mu_d}
   (j\mu_j 1,-\mu_d | I_{\mathrm B}\mu_{\mathrm B}) (-)^{\mu_d}\right. \nonumber\\
  &\times\sum_{m \mu_N} \left(lm\frac{1}{2}\mu_N | j \mu_j \right)
   \left. \bar{T}_{m,\mu_d' \mu_N \mu_d \mu_0}^{(pd)}
   \right|^2,
  \label{eq:dsigpd}
\end{align}
where
\begin{equation}
  C_0
  =\frac{{\cal M}_i {\cal M}_f}{(2\pi\hbar^2)^2}\frac{v_\beta}{v_\alpha}
  \label{eq:c0}
\end{equation}
with ${\cal M}_i$ (${\cal M}_f$) being the reduced energy of the $p$-A ($d$-B) system and $v_\alpha$ ($v_\beta$) their relative velocity.
Here, we make the following average prescription for $\mu_d$ and $\mu_N$ as in Ref.~\cite{Wakasa:2017rsk}:
\begin{eqnarray}
\tilde{t}^{*(pd)}_{\bm{\kappa}' \mu'_d \mu'_N, \bm{\kappa} \mu_0 \mu''_d}
\tilde{t}^{(pd)}_{\bm{\kappa}' \mu'_d \mu_N, \bm{\kappa} \mu_0 \mu_d}
&\approx& \frac{1}{6}\sum_{\bar{\mu}_N \bar{\mu}_d}
\left|
\tilde{t}^{(pd)}_{\bm{\kappa}' \mu'_d \bar{\mu}_N, \bm{\kappa} \mu_0 \bar{\mu}_d}
\right|^2 \nonumber \\
&&\times \delta_{\mu'_N \mu_N} \delta_{\mu''_d \mu_d}.
\end{eqnarray}
Then Eq.~\eqref{eq:dsigpd} reads
\begin{equation}
  \frac{d\sigma_{pd}}{d\Omega}
  =\frac{C_0}{6}S_{p}S_{d}\frac{2I_{\rm B}+1}{2l+1} \sum_m \frac{1}{6} \sum_{\mu'_d \bar{\mu}_N \mu_0 \bar{\mu}_d}
  \left| \bar{T}_{m,\mu_d' \bar{\mu}_N \bar{\mu}_d \mu_0}^{(pd)} \right|^2.
  \label{eq:dsigpd2}
\end{equation}
Using Eq.~\eqref{eq:barTamp3}, one finds
\begin{equation}
  \frac{d\sigma_{pd}}{d\Omega}
  =\bar{C}_0 \sum_m \left| \int d{\bm k}_d \, {\cal T}_{nljm}({\bm k}_d) \right|^2,
  \label{eq:dsigpd3}
\end{equation}
where
\begin{equation}
  \bar{C}_0 \equiv \frac{C_0}{6}S_{p}S_{d}\frac{2I_{\rm B}+1}{2l+1} \frac{(2\pi \hbar^2)^2}{{\cal M}^2_{pd}} \frac{1}{(2\pi)^6},
  \label{eq:c0bar}
\end{equation}
\begin{eqnarray}
  {\cal T}_{nljm}({\bm k}_d)
  &\equiv&
    \tilde{\psi}^{(d)}(\bm{k}_{d})
    \sqrt{\left(\frac{d \sigma_{pd}}{d\Omega} \right)_{\bm{\kappa', \kappa} } }
    \nonumber \\
    &&\times
    \int \chi^{(-)*}_{\bm{K}_{\rm f}} (\bm{R})
    \chi^{(+)}_{\bm{K}_{\rm i}}(\bm{R} )\psi^{(p)*}_{nljm}(\bm{R})
    e^{i \bm{k}_d \cdot \bm{R}}d{\bm R} \nonumber \\
  \label{eq:tmd}
\end{eqnarray}
with
\begin{equation}
  \left(\frac{d \sigma_{pd}}{d\Omega} \right)_{\bm{\kappa', \kappa} }
  =\frac{{\cal M}^2_{pd}}{(2\pi \hbar^2)^2}\frac{1}{6}\sum_{\mu'_d \bar{\mu}_N \mu_0 \bar{\mu}_d}
  \left| \tilde{t}^{(pd)}_{\bm{\kappa}' \mu'_d \bar{\mu}_N, \bm{\kappa} \mu_0 \bar{\mu}_d} \right|^2
  \label{eq:dsigpd4}
\end{equation}
and ${\cal M}_{pd}$ being the reduced energy of the $p$-$d$ system. In deriving Eq.~\eqref{eq:dsigpd3}, we have assumed
\begin{align}
   & \tilde{t}^{(pd)*}_{\bm{\kappa}'(\bm{k}'_d) \mu'_d \bar{\mu}_N, \bm{\kappa}(\bm{k}'_d) \mu_0 \bar{\mu}_d}
    \tilde{t}^{(pd)}_{\bm{\kappa}'(\bm{k}_d) \mu'_d \bar{\mu}_N, \bm{\kappa}(\bm{k}_d) \mu_0 \bar{\mu}_d} \nonumber \\
   & \approx \left|
    \tilde{t}^{(pd)*}_{\bm{\kappa}'(\bm{k}'_d) \mu'_d \bar{\mu}_N, \bm{\kappa}(\bm{k}'_d) \mu_0 \bar{\mu}_d}
    \tilde{t}^{(pd)}_{\bm{\kappa}'(\bm{k}_d) \mu'_d \bar{\mu}_N, \bm{\kappa}(\bm{k}_d) \mu_0 \bar{\mu}_d}
    \right|,
    \label{eq:tmatphase}
\end{align}
where the $\bm{k}_d$ dependence of $\bm{\kappa}'$ and $\bm{\kappa}$ is explicitly shown; the phases of $\tilde{t}^{(pd)}$ corresponding to different $\bm{k}_d$ are assumed to cancel out. According to Ref.~\cite{Yoshida:2024bbl}, this approximation is valid when the incident energy of the elementary process does not change significantly in the range of the integration. We will return to this point in Sec.~\ref{sec:origin}. 

The transition matrix distribution ${\cal T}_{nljm}$ in Eq.~\eqref{eq:tmd} indicates how the deuteron c.m. momentum contributes to the reaction process in terms of the momentum matching. This becomes more transparent if we consider the PW limit:
\begin{eqnarray}
  {\cal T}_{nljm}^{\rm PW}({\bm k}_d)
  &\equiv&
    \tilde{\psi}^{(d)}(\bm{k}_{d})
    \sqrt{\left(\frac{d \sigma_{pd}}{d\Omega} \right)_{\bm{\kappa', \kappa} } }
    \tilde{\psi}^{(p)*}_{nljm}(\bm{q}+\bm{k}_{d}) \nonumber \\
    &=&\sqrt{4\pi}\tilde{\varphi}^{(d)}(k_{d})
    \sqrt{\left(\frac{d \sigma_{pd}}{d\Omega} \right)_{\bm{\kappa', \kappa} } } \nonumber \\
    &&\times
    4\pi \tilde{\varphi}^{(p)}_{nlj}(|\bm{q}+\bm{k}_{d}|) i^{-l} Y^*_{lm}(\widehat{\bm{q}+\bm{k}_{d}}),
  \label{eq:tmd2}
\end{eqnarray}
where $\tilde{\varphi}^{(d)}$ and $\tilde{\varphi}^{(p)}_{nlj}$ are the radial parts of  $\tilde{\psi}^{(d)}$ and $\tilde{\psi}^{(p)}_{nljm}$, respectively, and ${\bm q}\equiv {\bm K}_{\rm i}-{\bm K}_{\rm f}$ is the momentum transfer. One can see from Eq.~\eqref{eq:tmd2} that ${\bm q}$ is shared by the momenta of $d$ in A and $p$ in B. This makes it easier to satisfy the momentum matching. On the other hand, in the standard PWBA framework, a neutron having the momentum of ${\bm q}$ in A is picked up.

\section{Results and discussion}
\label{sec:result}
\subsection{Numerical inputs}
This subsection details the numerical inputs for the DWIA formalism, which is the main focus of our theoretical development. 
The inputs specific to the conventional DWBA calculation will be discussed in Sec.~\ref{sec:comparison}, where a direct comparison between the two models is performed.

We adopt the EDAI parameter set of the Dirac phenomenology for nucleon optical potentials~\cite{Hama90,Cooper93}; for deuteron, we folded the proton and neutron potentials with $\phi^{(d)}({\bm r})$ calculated with the one-range Gaussian interaction~\cite{Ohm70}.
For the Coulomb potential for the deuteron distorted wave, a uniformly charged sphere with the reduced radius of 1.41~fm is considered.
As mentioned, we ignore the spin-orbit distorting potentials in the calculation of the distorted waves. To remedy this ignorance effectively, we slightly modified the central parts to reproduce the elastic cross sections calculated with both the original central and spin-orbit interactions.
In addition, to take into account the non-local nature of distorting potentials, we multiply each distorted wave by the Darwin factor for the case of the proton~\cite{Hama90, Arnold81} and the Perey factor with the nonlocality rage of 0.54 fm for the case of the deuteron~\cite{Perey62}.

\begin{figure}[htp]
  \includegraphics[width=7cm]{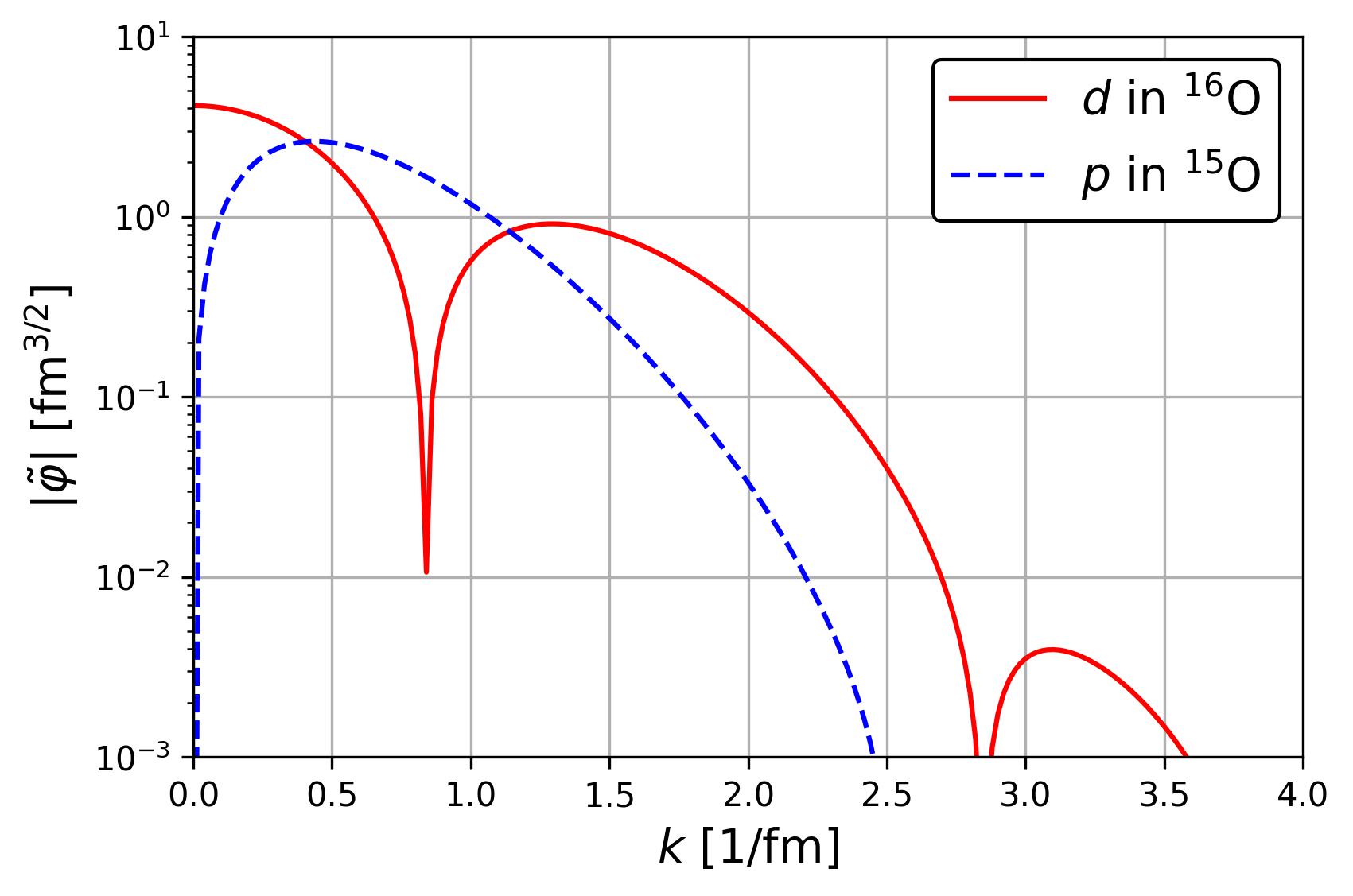}
  \caption{The momentum distributions of the deuteron (red solid line) and proton (blue dashed line).}
  \label{fig:2}
  \end{figure}

For the proton s.p. wave function, we use the Woods-Saxon potential proposed by Bohr and Mottelson~\cite{Bohr69}; we consider the 0${\rm p}_{1/2}$ orbital, hence $(n,l,j)=(0,1,1/2)$. The deuteron s.p. wave function is calculated with the Woods-Saxon potential with the reduced radius parameter 1.41~fm and the diffuseness parameter 0.65 fm~\cite{Samanta:1982xk}; we set the number of node to 1. The depths of the central parts are adjusted to reproduce the corresponding separation energies.
The radial parts of the deuteron and proton momentum distributions are shown in Fig.~$\ref{fig:2}$.

The $p$-$d$ elastic scattering cross-section data are taken from Refs.~\cite{
Davison1963, Kim1964, KURODA1964, HINTERBERGER1968, BUNKER1968, Cahill1971, Booth1971, SHIMIZU1982, Sagara1994, 
Sekiguchi2002, Hatanaka2002, Ermisch2005}, and the Lagrange interpolation is performed with respect to the scattering angle and energy, following the procedure outlined in Ref.~\cite{Yoshida:2024bbl}.
To describe the transformation of the $p$-$d$ transition matrix from the $p$-$d$ c.m. frame to the $p$-$^{16}$O c.m. frame, the M{\o}ller factor is taken into account~\cite{Moller1945, Kerman59}.
For the off-the-energy-shell cross section of the elementary process, one can approximate it to the on-the-energy-shell cross section with conventional prescriptions. 
Here, we adopt the final-state prescription, in which the momentum of the initial channel is taken to be $\bm{\kappa} \approx \kappa' \bm{\hat{\kappa} }$.

\subsection{The transfer cross section for $^{16}$O$(p,d)^{15}$O}
\begin{figure}[htp]
  \includegraphics[width=7cm]{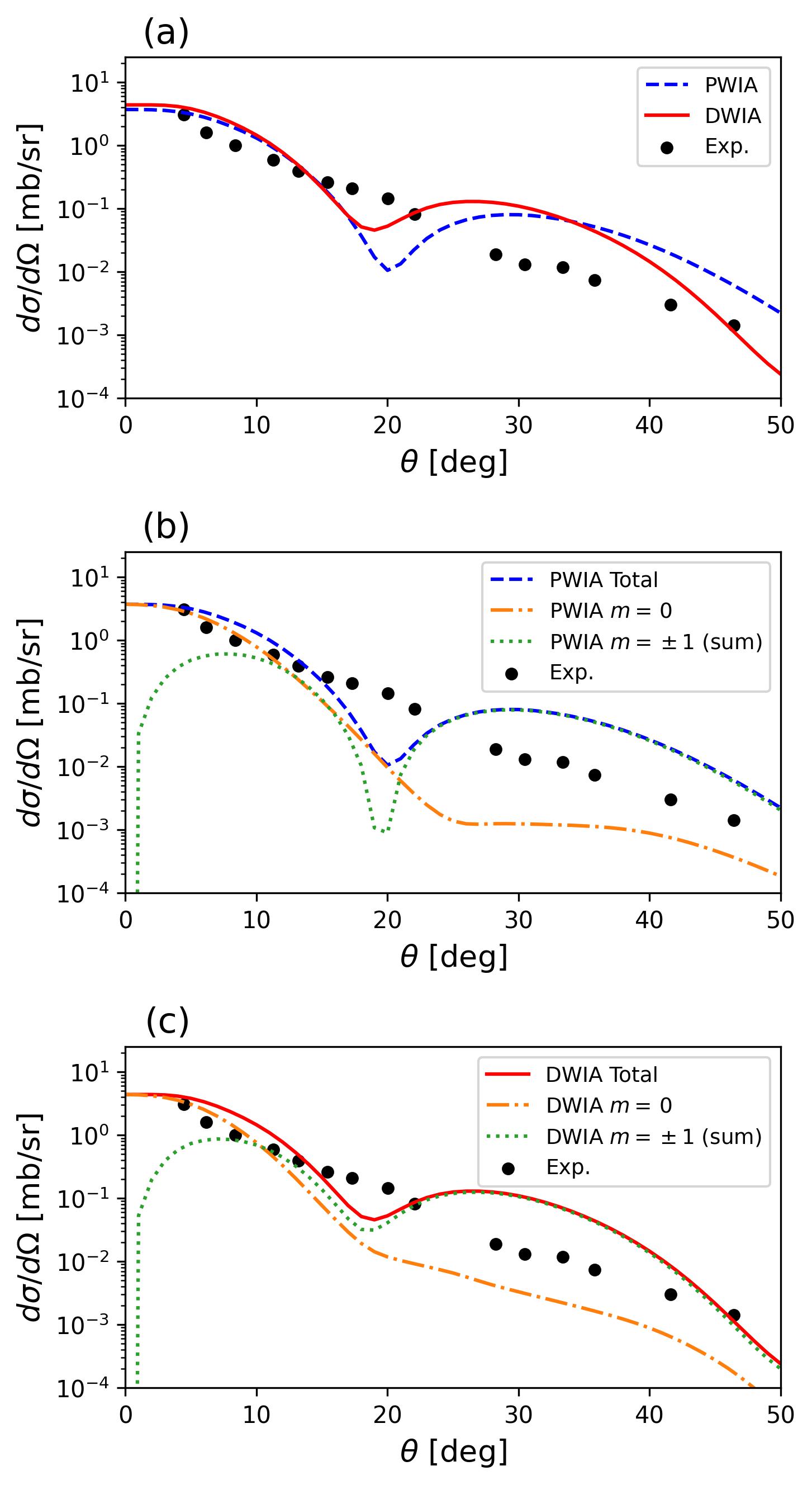}
  \caption{The transfer cross section of $^{16}$O$(p,d)^{15}$O at 200~MeV as a function of the scattering angle $\theta$. (a) Comparison between the results with DWIA (red solid line) and PWIA (blue dashed line). (b) The PWIA result (red solid) is decomposed into the $m=0$ (orange dash-dotted line) component and the sum of the $m=\pm 1$ (green dotted line) components. (c) Same as (b) but for the DWIA result.}
  \label{fig:3}
\end{figure}

Using the framework developed in the present work, we calculated the cross sections of the transfer reaction $^{16}$O$(p,d)^{15}$O at 200~MeV with PWIA and DWIA, as shown in Fig.~\ref{fig:3}. 
In Fig.~\ref{fig:3}(a), the experimental data~\cite{Abe89} and calculated results are presented as a function of the deuteron emitting angle $\theta$ in the c.m. frame.
The theoretical results are scaled to match the experimental data in the region $4^{\circ}<\theta<15^{\circ}$. 
The scaling factors for PWIA and DWIA results are $20$ and $100$, respectively, which include the product of $S_{p}$ and $S_{d}$. 
The difference in the PWIA and DWIA scaling factors indicates that the absorption effect makes the cross section smaller by about a factor of five.
Although the product $S_p S_d$, which corresponds to states with the excited $^{14}$N core, is not well determined, these large scaling factors are somewhat unexpected.
We will discuss this point in more detail in the following subsections.

In Fig.~\ref{fig:3}(b), the PWIA cross section, which is the same as in Fig.~\ref{fig:3}(a), is shown along with the contributions from the $m=0$ and the sum of the $m=\pm 1$ components; the results obtained with DWIA are shown in Fig.~\ref{fig:3}(c). It should be noted that the $m=1$ and $m=-1$ components are the same.
One can see that for both PWIA and DWIA, the contribution of $m=0$ is dominant at small angles near $\theta=0^{\circ}$, whereas the contribution of $m=\pm 1$ becomes more significant at somewhat larger angles around $\theta=40^{\circ}$.

\subsection{Momentum matching}

As mentioned, the momentum matching is crucial for transfer reactions. Although the present framework is expected to ease the momentum-matching condition, the unexpectedly large scaling factors found, i.e., the cross sections being too small to explain the experimental data, will indicate that the above expectations may not hold. To make the situation clearer, we analyze the transition matrix distributions, those in the PW limit given by Eq.~\eqref{eq:tmd2}, for $\theta=0^\circ$ and $40^\circ$. It turns out that the features of the results shown below remain when the distortion is included. We follow the Madison convention for the kinematics of the reaction particles.

\begin{figure}[htp]
  \includegraphics[width=7cm]{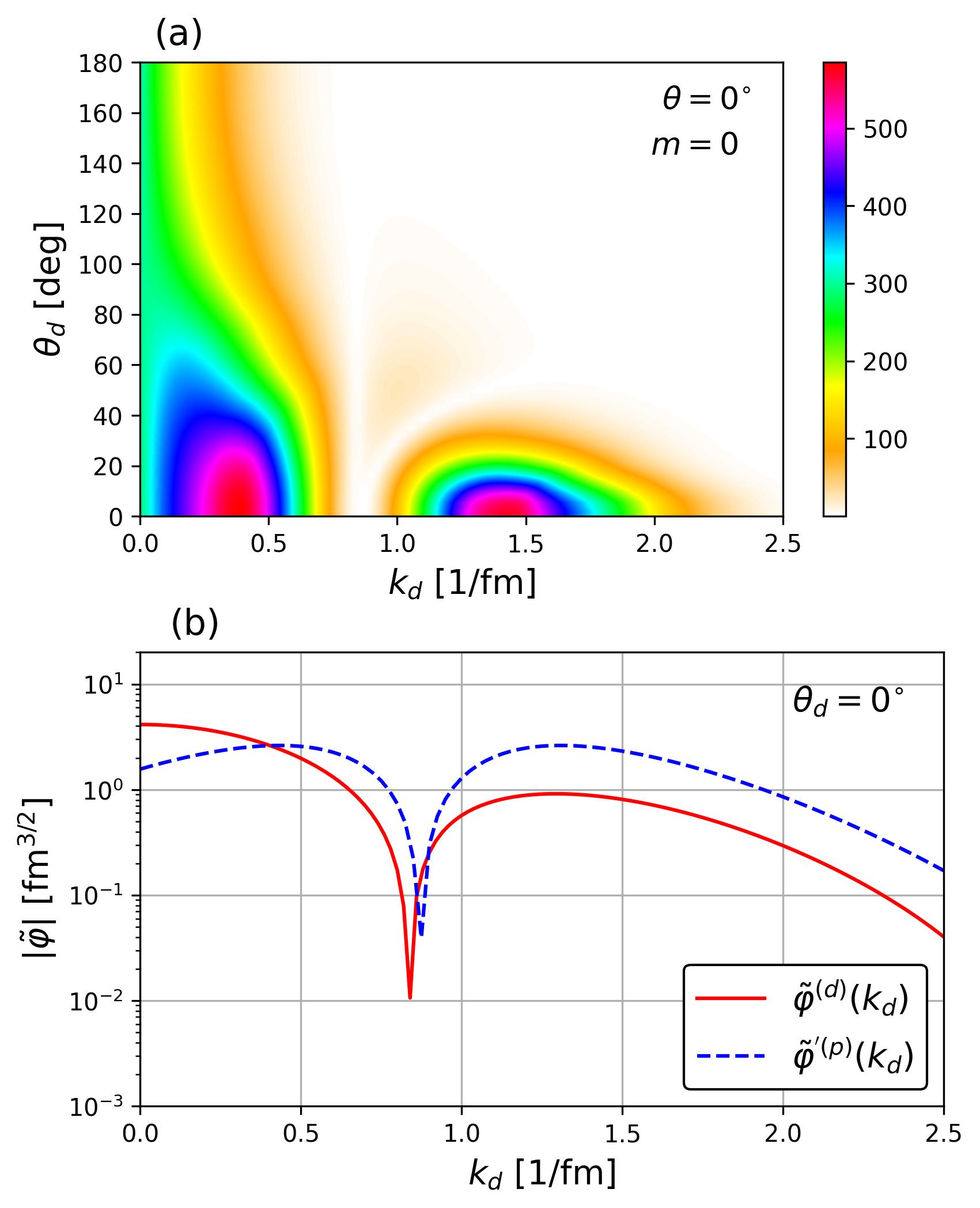}
  \caption{(a) $\left|\mathcal{T}_{01,1/2,0}^{\rm PW}\right|$ at $\theta=0^{\circ}$ as a function of $k_d$ and $\theta_d$. (b) The deuteron (red solid line) and proton (blue dashed line) momentum distributions along $\theta_d=0^{\circ}$.}
  \label{fig:4}
\end{figure}

One sees from Eq.~\eqref{eq:tmd2} that at $\theta=0^\circ$, ${\cal T}_{nljm}^{\rm PW}$ has no contribution to the cross section unless $m=0$, and is purely imaginary for $l=1$. In Fig.~\ref{fig:4}(a), $\left| \mathcal{T}_{01,1/2,0}^{\rm PW} \right|$ is shown as a function of $k_d$ and $\theta_d$; it does not depend on the azimuthal angle $\phi_d$ of $\bm{k}_d$. It has two local maxima near $(k_d,\theta_d) = (0.3~{\rm fm}^{-1}, 0^\circ)$ and $(1.4~{\rm fm}^{-1}, 0^\circ)$. These maxima reflect the matching of the deuteron and proton momentum distributions for the given momentum transfer.

In Fig.~\ref{fig:4}(b), the $k_d$ dependence of $\tilde{\varphi}^{(d)}$ (the red solid line) and $\tilde{\varphi}^{(p)}_{01,1/2}$ (the blue dashed line) at $\theta_d=0^\circ$ is shown; the former is the same as in Fig.~\ref{fig:2}. As shown, $\tilde{\varphi}^{(d)}$ has a node around $k_d^{(0)} \equiv 0.8~{\rm fm}^{-1}$. Although it cannot directly be seen from Fig.~\ref{fig:2} or Fig.~\ref{fig:4}(b), $\tilde{\varphi}^{(d)}$ changes its sign at $k_d = k_d^{(0)}$. The behavior of $\tilde{\varphi}^{(p)}_{01,1/2}$ is understood by the $k_d$ dependence of its argument $k_p \equiv |\bm{q}+\bm{k}_{d}|$. At  $\theta=0^\circ$, $q\sim 0.9$~fm$^{-1}$ and ${\bm q}$ is antiparallel to the $z$-axis. For $0\le k_d \le q$, $k_p$ becomes smaller as $k_d$ increases, with keeping ${\bm k}_p$ antiparallel to the $z$-axis, and $\tilde{\varphi}^{(p)}_{01,1/2}=0$ at $k_d=q$. Then, as $k_d$ increases further, $k_p$ becomes larger with ${\bm k}_p$ being parallel to the $z$-axis. Because of the property of the spherical harmonics $Y_{1m}$, $\tilde{\psi}^{(p)}_{01,1/2,0}$ changes its sign at $k_d=q$. Therefore, the product $\tilde{\psi}^{(d)} \, \tilde{\psi}^{(p)}_{01,1/2,0}$ keeps its sign except in the very narrow region of $k_d^{(0)} < k_d < q$. This implies that the momentum sharing by $d$ in $^{16}$O and $p$ in $^{15}$O works well.

\begin{figure}[htp]
  \includegraphics[width=7cm]{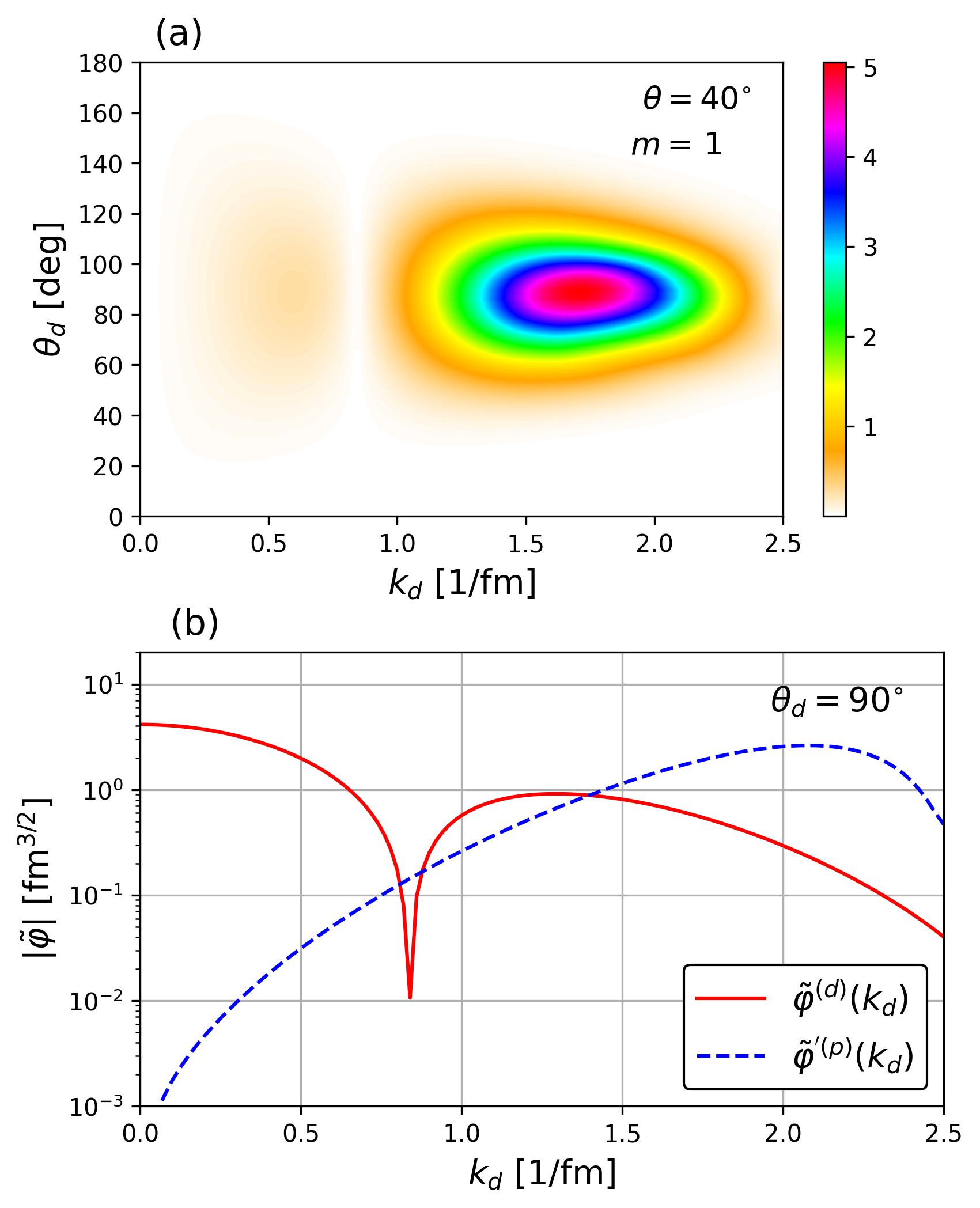}
  \caption{Same as Fig.\ref{fig:4} but for $\theta=40^{\circ}$. See the text for details.}
  \label{fig:5}
\end{figure}
A similar analysis is carried out for $\theta=40^\circ$ as shown in Fig.~\ref{fig:5}. As discussed above, the $m= \pm 1$ component is dominant and we focus on 
$\mathcal{T}_{01,1/2,1}^{\rm PW}$; it has both the real and imaginary parts but the former turns out to negligibly be small. For the $\phi_d$ dependence, it was found that $\mathcal{T}_{01,1/2,1}^{\rm PW}$ mainly distributes around $\phi_d=0$. Thus, we plot $\left|{\rm Im} (1/\pi) \int_{-\pi/2}^{\pi/2}\mathcal{T}_{01,1/2,1}^{\rm PW}(\bm{k}_d)\, d\phi_d \right|$ in Fig~\ref{fig:5}(a). One sees that it is well concentrated around $(k_d,\theta_d) = (1.7~{\rm fm}^{-1}, 90^\circ)$. The $k_d$ dependence of $\tilde{\varphi}^{(d)}$ and $\tilde{\varphi}^{(p)}_{01,1/2}$ is shown in Fig~\ref{fig:5}(b) in the same way as in Fig~\ref{fig:4}(b) but along $\theta_d=90^\circ$. At $\theta=40^\circ$, $q\sim 2.5$~fm$^{-1}$ and $\bm{q}$ is antiparallel to the $x$-axis (in the Madison convention). One sees from Fig~\ref{fig:5}(b) that an optimal momentum sharing is realized with $k_d\sim 1.7$~fm$^{-1}$ and $k_p\sim 0.8$~fm$^{-1}$. In the standard DWBA, the momentum transfer of 2.5~fm$^{-1}$ must be carried by the neutron momentum in $^{16}$O. 

To summarize the results in this subsection, contrary to expectations based on the anomalously large scaling factors found in Fig.~\ref{fig:3}(a), the sharing of the large momentum transfer by $d$ in $^{16}$O and $p$ in $^{15}$O works well. To check this phenomenologically, we changed the Woods-Saxon parameters of the deuteron and proton binding potentials. The transfer cross sections varied to some extent but the scaling factors remained in the same orders. We will discuss other possible sources of this issue in the next subsection.

\subsection{Possible sources of anomalously large scaling factors}
\label{sec:origin}

First, let us consider Eq.~\eqref{eq:tmatphase}, which is used to represent the transition matrix of the transfer reaction with the cross section of the elementary process. If we used the $p$-$d$ transition amplitude, the angular momentum algebra became much more complicated. As mentioned, this approximation holds well when the energy $E_{pd}$ of the elementary process does not change significantly in the range of the integration~\cite{Yoshida:2024bbl}. We have checked the range of $E_{pd}$ relevant to the integration over $\bm{k}_d$ and found that it varies from a few MeV to several hundreds of MeV. Although quite restricted regions of $\bm{k}_d$ are important (see Figs.~\ref{fig:4}(a) and \ref{fig:5}(a)) and the actual range of $\bm{k}_d$ to be considered will not be so wide, this $E_{pd}$ dependence can violate Eq.~\eqref{eq:tmatphase}. Taking the $p$-$d$ transition amplitude into account will be crucial to conclude the absolute value of the $(p,d)$ cross section calculated with DWIA.

Another origin of the problem can be the deuteron bound state wave function. 
Although we consider the 1${\rm s}$ state for the deuteron in $^{16}$O, the 0${\rm d}$ orbital cannot be excluded. According to Ref.~\cite{Samanta:1982xk}, the 0${\rm d}$ state contribution seems to be nonsignificant because the $(p,pd)$ knockout cross section has a single peak, which indicates the ${\rm s}$-wave state of the deuteron in $^{16}$O. However, there will be a possibility that the role of the 0${\rm d}$ state in the transfer reaction is different from that in the knockout reaction. It may be essential to calculate the contribution of the 0${\rm d}$ orbital to appropriately explain the experimental data.

\subsection{DWIA vs. DWBA}
\label{sec:comparison}
In this subsection, we perform a direct comparison between the transfer cross sections calculated with the DWIA and DWBA formalisms.

Before presenting the results, we summarize the numerical inputs for the PWBA and DWBA calculations. 
The neutron s.p. wave function was generated using the Bohr-Mottelson potential as in the DWIA case. 
The strength of the zero-range interaction, $D_0$, was set to $125~\text{MeV}\cdot\text{fm}^{3/2}$, which is consistent with the value used in Ref.~\cite{Ohm70}. 
For the PWBA and DWBA calculations, the neutron spectroscopic factor, $S_{nlj}$, is set to unity. 
Similarly, for the PWIA and DWIA calculations, the proton ($S_p$) and deuteron ($S_d$) spectroscopic factors are also set to unity. 
For this comparison, no additional scaling factors are included in the calculations.

\begin{figure}[htp]
  \includegraphics[width=7cm]{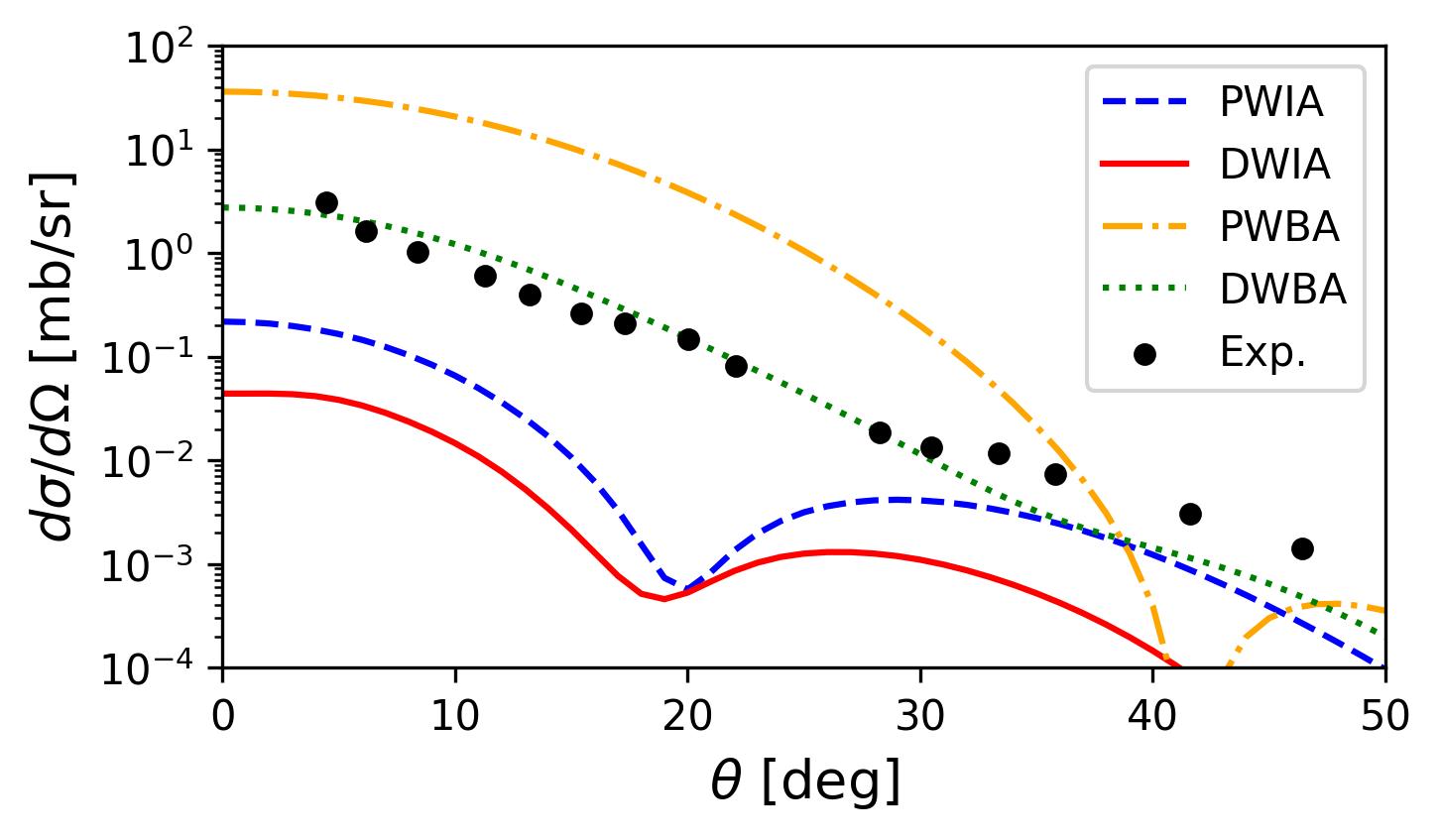}
  \caption{The transfer cross section of $^{16}$O$(p,d)^{15}$O at 200 MeV obtained with the DWIA (red solid line) and DWBA (green dotted line) calculations without scaling factors. The corresponding PWIA and PWBA results are also shown by the blue-dashed and orange-dash-dotted lines, respectively.}
  \label{fig:6}
\end{figure}
Figure~\ref{fig:6} shows the results of this comparison. 
It displays the DWIA result (red solid line) and the DWBA result (green dotted line), together with the corresponding PWIA (blue dashed line) and PWBA (orange dash-dotted line) results. 
The experimental data are the same as those shown previously. 
From the figure, one sees that the PWIA and DWIA cross sections are smaller than the experimental data by factors of about 20 and 100, respectively, whereas the PWBA result is larger than the data by roughly a factor of 10. 
In striking contrast, the DWBA result reproduces the experimental data well. 
Considering that the spectroscopic factor $S_{nlj}$ is expected to be in the range of 1 to 2~\cite{Abe89,Bachelier:1969sfv,Towner:1969tda}, the DWBA calculation successfully describes the experimental data without the need for any unnatural scaling factors.

The results in Fig.~\ref{fig:6} are very interesting. 
The DWBA results seem to reproduce the experimental data better than the DWIA results. 
However, this raises the question: why? In some sense, such a situation might even be regarded as having driven us further away from resolving the problem.
One possible explanation for this difference is that, in DWBA, the initial $^{16}$O nucleus is treated as a system of $^{15}$O and a neutron, whereas in DWIA it is considered as a system of $^{14}$N and a deuteron. 
This different treatment of the nuclear structure may lead to a significant difference in the cross section.

Furthermore, the apparent superiority of DWBA revealed in this comparison may depend on the specific nucleus and reaction energy considered; in other cases, the relative performance of DWIA and DWBA could be reversed.
To clarify this issue, further studies on transfer reactions with available experimental data are necessary. Although some works based on DWBA have already been carried out~\cite{Pang:2014kha,Xu:2024ohi}, systematic comparisons between DWIA and DWBA are essential for deepening our understanding of nuclear structure and transfer reactions.

It is also worth noting the effect of distortion on the angular distributions.
The angular shapes of the DWIA and PWIA calculations are quite similar, and both qualitatively follow the experimental trend.
In the Born approximation, however, the inclusion of distortion drastically changes the shape of the distribution; the PWBA result does not seem to match the data, whereas the DWBA result shows good agreement with the experiment.
\section{Summary and conclusions}
\label{sec:summary}
In this work, we have investigated the validity of DWIA for describing the ($p,d$) transfer reaction at intermediate energies by performing a comparative study with DWBA. 
We applied both formalisms to the $^{16}$O($p,d$){}$^{15}$O reaction at an incident energy of 200~MeV. 
The DWBA calculation successfully reproduced the experimental data in both the shape of the angular distribution and the absolute magnitude of the cross section. 
In contrast, while the DWIA result showed a similar trend in the angular distribution, it severely underestimated the magnitude of the cross section by a factor of about 100.

We also investigated the transition matrix distribution of Eq.~\eqref{eq:tmd2} on the $k_d$-$\theta_d$ plot; for clear interpretation we analyzed the results with PWIA. 
At each of $0^\circ$ and $40^\circ$, the relatively large momentum transfer is found to be shared by the deuteron in $^{16}$O and proton in $^{15}$O. 
As a result, a very high momentum of the bound particle in the nucleus is not needed, on the contrary to the standard DWBA picture. 
These features are found to remain when we use DWIA.

This result raises a crucial open question as to why the transfer reaction at this intermediate energy is not well described by the present DWIA framework, unlike knockout reactions. 
A potential reason for the significant discrepancy in the DWIA calculation could be the phase cancellation assumption for the $p$-$d$ transition amplitude used in Eq.~\eqref{eq:tmatphase}. 
A more sophisticated treatment of the $p$-$d$ amplitude, as well as the inclusion of contributions from the deuteron in the $0d$ orbital in $^{16}$O, will be necessary to draw a conclusion on the applicability of DWIA to this reaction.

In conclusion, our findings suggest that for the $^{16}$O($p,d$){}$^{15}$O reaction at 200~MeV, the DWBA provides a more accurate description of the experimental data. 
This motivates the need for more systematic investigations to clarify the conditions, such as higher incident energies, larger momentum transfers, or different nuclear targets (e.g. Lithium isotopes), under which the DWIA might become a better approach for transfer reactions. 
Such comprehensive studies are essential to delineate the applicability of these two fundamental reaction mechanisms.
\section*{Acknowledgments}
We thank H.~J.~Ong and S.~Terashima for fruitful discussions.
This work is supported in part by Grants-in-Aid for Scientific Research from the JSPS (Grants No. JP20K14475, No. JP21H04975, and No. JP25K17400) and JST ERATO Grant No. JPMJER2304, Japan. S.-I.~S. has been supported in part by the NRF grants (NRF-2018R1A5A1025563, NRF-2022R1A2C1003964, and NRF-2022K2A9A1A06091761).

\appendix
\end{document}